%% file: main.tex
\documentclass[sigconf,screen]{acmart}
\input{macro}
\AtBeginDocument{%
  }

\acmConference[FSE Companion ’25]{}{June 23--275,
  2025}{Trondheim, Norway}
\acmBooktitle{Companion Proceedings of the 33rd ACM Symposium on the Foundations of Software Engineering (FSE '25), June 23--27, 2025, Trondheim, Norway}

\begin{document}

\title{A Tool for In-depth Analysis of Code Execution Reasoning of Large Language Models}


\author{Changshu Liu}
\affiliation{%
  \institution{University of Illinois Urbana Champaign}
  \city{Urbana}
  \country{USA}}
\email{cl144@illinois.edu}

\author{Reyhaneh Jabbarvand}
\affiliation{%
  \institution{University of Illinois Urbana Champaign}
  \city{Urbana}
  \country{USA}}
\email{reyhaneh@illinois.edu}

\begin{abstract}
Code Executing Reasoning is becoming a new non-functional metric that assesses the ability of large language models (LLMs) in programming tasks. State-of-the-art frameworks (CodeMind or REval) and benchmarks (CruxEval) usually focus on LLM's prediction of a given code's input/output or intermediate variable states/values on limited programs. However, there is no tool for more in-depth analysis of the results. Without such a tool, the observations about LLM's code execution reasoning cannot be generalized to more datasets, preventing the research community and practitioners from devising the next generation of LLMs with better code execution reasoning abilities. This paper introduces \name, a series of tools and heuristics to analyze the result of code execution reasoning frameworks to understand better the impact of code properties in the studied benchmarks on the code execution reasoning. With such tooling, analysis can be generalized to code with similar properties without the urgent need to design more benchmarks, which is a cumbersome effort. 

The implementation of \name is publicly available, and it currently assesses the impact of different (1) program constructs, (2) program complexities, (3) dynamic programming properties such as recursion length, and (4) variable types on code execution reasoning abilities of LLMs. Evaluation of \name on four programming benchmarks ($1450$ Python programs) and six LLMs from three existing code execution reasoning techniques shows its effectiveness and practicality in providing important insights, highlighting the strengths and limitations of LLMs concerning code execution reasoning. 

\end{abstract}

\maketitle

\input{sections/introduction}

\input{sections/related-work}
\input{sections/approach}

\input{sections/evaluation}

\input{sections/conslusion}

\bibliographystyle{ACM-Reference-Format}
\bibliography{main}


\end{document}

%% file: macro.tex
\usepackage{graphicx}
\usepackage{amsmath} 
\usepackage{stmaryrd}
\usepackage{xspace}
\usepackage{cleveref}
\usepackage{multirow}
\usepackage{nth}
\usepackage{wrapfig}
\usepackage[font=small,labelfont=bf]{caption}
\usepackage{subcaption}
\usepackage{enumitem}
\usepackage{bm}
\usepackage{soul,color}
\usepackage{listings}
\usepackage{float}
\usepackage{subcaption}
\usepackage{verbatim}

\newcommand{\name}{ExeRScope\xspace}

\newcommand{\gptf}{GPT-4\xspace}

\newcommand{\gemini}{Gemini-1.5-Pro\xspace}
\newcommand{\deepseek}{DeepSeekCoder\xspace}
\newcommand{\codellama}{CodeLlama\xspace}

\newcommand{\sem}{SemCoder-S\xspace}
\newcommand{\heval}{HumanEval\xspace}

\newcommand{\avatar}{Avatar\xspace}
\newcommand{\crux}{CRUXEval\xspace}
\newcommand{\ceval}{ClassEval\xspace}
\newcommand{\codemind}{CODEMIND\xspace}
\newcommand{\starT}{StarCoder\,2\xspace}
\newcommand{\reval}{\textsc{REval}\xspace}

\definecolor{codegreen}{rgb}{0,0.6,0}
\definecolor{codegray}{rgb}{0.5,0.5,0.5}
\definecolor{codepurple}{rgb}{0.58,0,0.82}
\definecolor{backcolour}{rgb}{0.95,0.95,0.92}

\lstdefinestyle{stylewithcommentPy}{
language=Python,
 basicstyle=\linespread{0.9}\footnotesize,
     backgroundcolor=\color{backcolour}, commentstyle=\color{codegreen},
  keywordstyle=\color{magenta},
  numberstyle=\tiny\color{codegray},
  basicstyle=\scriptsize\ttfamily,
  escapechar=|,
  stringstyle=\color{codepurple}, xleftmargin=0cm,
    frame=tlbr, framesep=0.1cm, framerule=0pt, numbers=none,
    breaklines=true,
    moredelim=**[is][\color{red}]{@}{@},
    moredelim=**[is][\color{black}]{`}{`},
    moredelim=[is][\color{red}\bfseries\underbar]{~}{~},
    moredelim={[is][\color{black}\bfseries]{@@}{@@}}
}

%% file: sections/introduction.tex
\section{Introduction}

Large Language Models (LLMs) have shown emerging abilities in code/test synthesis, bug/vulnerability detection, code translation, and program repair~\citep{roziere2023code,zhu2024deepseek,achiam2023gpt,lozhkov2024starcoder,reid2024gemini,yang2024qwen2,mishra2024granite}. The extent to which they can reason about code execution is becoming a new metric to assess their programming abilities and generalizing such abilities beyond crafted benchmarks. Three prominent frameworks for evaluating LLM's code execution reasoning are \crux~\cite{gu2024cruxeval}, CodeMind~\cite{liu2023code}, and \reval~\cite{chen2024reasoning}. \crux is a benchmark of $800$ synthetic Python functions generated by LLMs and their corresponding input-output pairs. The dataset comes with two code execution reasoning tasks: Output Prediction, which requires the model to predict the output of the function given the input, and Input Prediction, which requires the model to predict the input given the output after the code execution. 

\codemind is another code reasoning framework that proposes two main code reasoning tasks: Execution Reasoning (ER), which assesses LLM's ability to predict execution output given provided inputs, and Specification Reasoning (SR), which evaluates LLM's understanding of the natural language specification and semi-formal specification in the form of test input/output. \reval~\cite{chen2024reasoning} takes one step further than two previous approaches in execution reasoning and evaluates LLMs using four runtime behavior prediction tasks: for given inputs and a statement in the program, it asks LLMs to predict (1) if the statement is covered during execution(Code Coverage Prediction), (2) variable values after the execution of it (Program State Prediction), (3) the following statement executed after it (Execution Path Prediction), and (4) final output (Output Prediction).

These frameworks suffer from the following limitations: they primarily focus on comparing predicted variable states and similar results with the collected ground truth while neglecting a deeper analysis of the factors influencing LLM performance in code execution reasoning. Examples of such deeper analyses are answering (1) how different code constructs (e.g., conditional statements and recursive constructs) affect LLMs' understanding of code; (2) will LLMs be confused about reasoning about the execution of longer loops (with more iterations); (3) are LLMs sensitive to particular variable types; (4) and what is the overall impact of code complexity levels on their performance. More importantly, claiming the generalizability of the observations beyond the evaluated benchmarks is theoretically impossible~\cite{zhang2024can,kang2024learning}. Even if one considers empirical evaluation as an upper bound for the capabilities of LLMs, evaluating several LLMs under different benchmarks can be time-consuming and require non-trivial resources. 

To advance the state of code execution reasoning evaluation of LLMs, we introduce \name (\underline{Exe}cution \underline{R}easoning \underline{Scope}), a tool that incorporates a program analysis to provide insights into the code execution reasoning capability of LLMs beyond value prediction. For a series of different programs and their corresponding code execution reasoning results (i.e., variable value predictions and whether the LLM correctly predicted the value or not), \name extracts specific properties (program constructs, code complexity, dynamic properties such as recursion length, and output type) and investigates how these factors vary in correct predictions versus incorrect ones. It then plots the results with appropriate visualization, aiding model developers in understanding the limitations of their LLMs better. 

\sloppy Utilizing \name on \emph{output prediction} results of six LLMs (\gptf, \gemini, \deepseek, \codellama, \sem , and \starT) under $1450$ Python programs from \emph{four} widely used programming benchmarks (\heval, \avatar, \ceval, and \crux) reveals that recursive and nested program constructs, higher program complexity, more loop iterations as well as non-primitive types negatively impact the reasoning ability of LLMs. Given that these properties are natural in real-world programs, these results emphasize the importance of devising benchmarks reflecting these complexities. Our evaluation of \name is based on output prediction since it is the only common execution reasoning task in existing frameworks. The current version also supports the analysis of Python programs since Python was the common programming language among the subject programs used in existing frameworks. However, \name can take any arbitrary program and its corresponding execution reasoning results in Java or Python and update the analysis.

%% file: sections/related-work.tex
\section{Related Work/Research Gap}

\begin{figure}[t]
    \centering
    \includegraphics[width=0.99\linewidth]{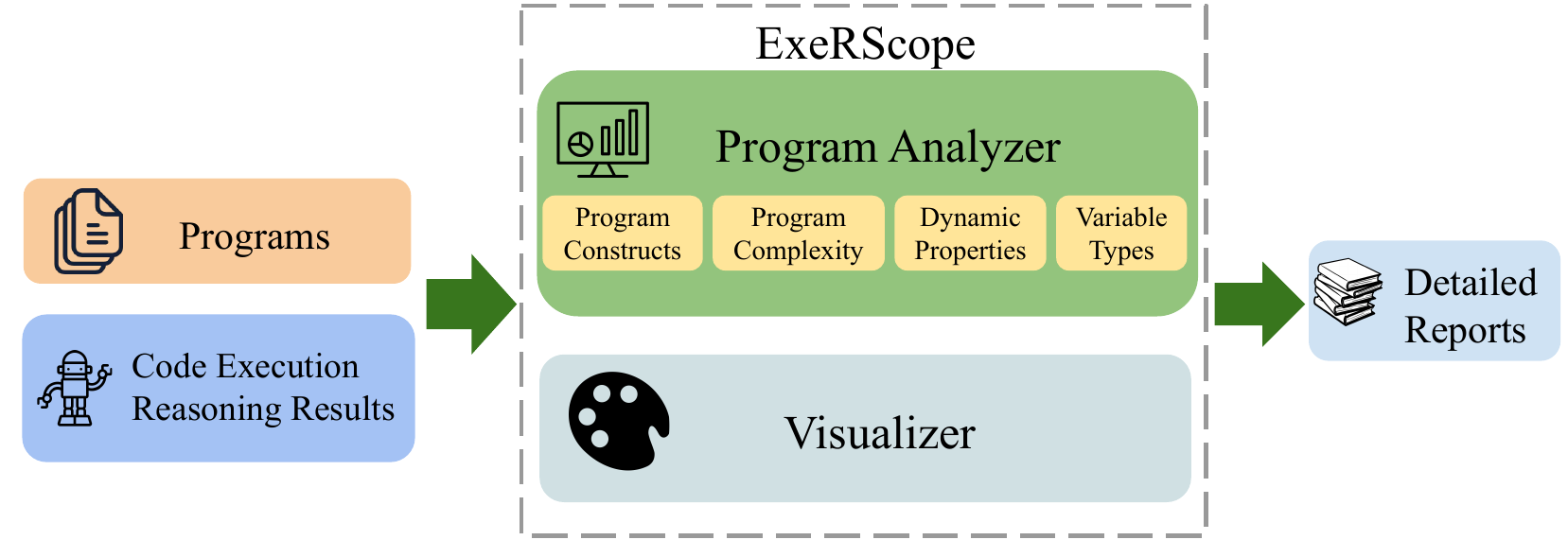}
    \caption{Overview of \name.
    }
    \label{fig:overview}
    \vspace{-15pt}
\end{figure}

A substantial amount of research has evaluated LLMs on reasoning tasks across various modalities \cite{deshpande2021rect,wu2023reasoning,miceli2023larger,bubeck2023sparks,wang2023mathcoder,imani2023mathprompter,luo2023wizardmath,huang2023lvlms,valmeekam2022large,min2023beyond}, such as natural language, visual data, mathematics, logic, and code. \name is more directly aligned with recent studies specifically addressing code reasoning ~\cite{la2024code,gu2024cruxeval,zhang2024transformer, chen2024reasoning}. To our knowledge, these frameworks only propose code reasoning tasks and evaluate LLMs based on some defined metrics rather than proposing a systematic way of profoundly analyzing the results. \emph{\name can be plugged into these code reasoning frameworks and complement them with additional insights from grounded programming language heuristics and sound program analysis techniques.}  

Another related line of research to \name is developing execution-aware Code LLMs that are pre-trained or instruction-tuned using execution information to perform programming tasks better. NeXT~\cite{ni2024next} trains LLMs to analyze execution traces and produce natural language explanations for reasoning about program run-time behavior. SemCoder \cite{ding2024semcoder} guides LLMs using operational semantics and natural language dialogues to execute the program step by step. These techniques rely on synthetic data to instruction-tune LLMs for code synthesis in general (SemCoder) or specific to program repair (NeXT). In designing such a dataset, they do not consider the factors that we will show impacting the code execution reasoning abilities of the LLMs (and very likely the overall programming performance). As a result, SemCoder, although shown to outperform bigger LLMs in code execution reasoning under \crux, suffers to generalize to longer loops or complex types in other studied datasets. \emph{These observations highlight the importance of tools such as \name to provide better insights into developing better code LLMs, from factors that should be considered in designing datasets to instruction-tuning objectives.} 

%% file: sections/approach.tex
\section{\name Tool}

Figure~\ref{fig:overview} provides an overview of \name. \name takes LLMs, subject programs, and code execution reasoning results as inputs and provides a detailed report about the performance of LLMs over the programs concerning code execution reasoning. \name comprises two components: \textbf{Program Analyzer}, which combines static and dynamic techniques to extract a series of properties from the programs, and \textbf{Visualizer}, which represents the analysis results with proper figures and generates reports including tables and visualized figures. The static analysis identifies different program constructs in each subject, measures the program complexity using multiple metrics, and determines the variable types. Additionally, the dynamic analysis executes the program using input data to determine the number of iterations in the presence of recurring constructs, e.g., how many times a for loop is executed or how many times a recursive function is called.

\subsection{Program Analysis}

The program analyzer component uses static and dynamic analysis to extract different information about the subject programs, namely program constructs in the implementation, code complexity, programming properties, and variable types, to incorporate into the analysis. We will explain the details of the analysis related to each property as follows.

\subsubsection{Program Constructs}
\label{sec:construct}
\name tags each subject program with at least one of the following labels: \textbf{Basic} (for programs with no specific programming constructs), \textbf{If} (for programs with if statement), \textbf{For} (for programs with for loop), \textbf{Match} (for programs with match-case construct), \textbf{Try} (for programs with exception handling construct), and \textbf{While} (for programs with while loop). The rationale here is that each program constitutes these programming constructs, sequentially or compound, regardless of how easy or complex it is and whether it is from the real world or synthetic. Therefore, analyzing the execution reasoning results concerning different programming constructs can be a generic way to identify their strengths and weaknesses to reason about them. The current version also complements the compound label observed in the studied subject, i.e., \textbf{Nested Ifs} and \textbf{Nested Loops}\footnote{We plan to augment the program analysis pipeline and include more labels in the next versions of \name, supporting more programs in other benchmarks, if used by code reasoning frameworks.}. We implement the programming construct tagging using Python's \emph{ast} module~\cite{ast}.

\subsubsection{Program Complexity} 
\label{subsec:complexity}
Another program property that can intuitively impact the code execution reasoning abilities of LLMs is program complexity. \name supports different code complexity metric measurements, including lines of code, cyclomatic complexity, and cognitive complexity. The higher the lines of code for a program, the more context the LLM should consider when reasoning about code execution, making it more vulnerable to drop code execution reasoning performance. Cyclomatic complexity measures the independent execution paths in the program control flow graph (CFG). Cyclomatic complexity (CC) can be measured as $CC=E-N+2P$, where $E$ and $N$ are the number of edges and nodes in the CFG, respectively, and $P$ is the number of methods when we measure CC in for class. For code execution reasoning, the higher the number of independent paths, the more likely it is challenging for the LLM to succeed by chance. Cognitive complexity is similar to cyclomatic complexity, with more emphasis on the nested constructs, which could seemingly be more challenging for the LLMs to reason about the program execution through them. 

While tools and libraries exist to measure cyclomatic and cognitive complexities (e.g., lizard~\cite{lizard}), they can only handle method-level complexity and fail to reveal class dependencies when multiple methods exist. So, instead, \name leverages PyCFG library to build CFGs and compute the complexity metrics. 

\subsubsection{Dynamic Properties}
During execution reasoning, even if it is a prediction of the final output value, LLMs are expected to simulate the program's execution given the inputs. Sometimes, the code syntax and static properties can be simple, but dynamic properties can be complex concerning the execution reasoning. For example, consider the code snippet in Listing~\ref{lst:python-recursive} implementing a simple recursive logic using a for loop. When \texttt{n} is ten, an LLM needs to keep track of only ten values of the \texttt{diff} list to reason about output values after loop terminations. However, this number increases for larger values of \texttt{n}, e.g., $100$, requiring more reasoning/memorization steps and challenging LLMs more. As a result, in addition to static properties of the programs, \name uses dynamic analysis to incorporate dynamic properties into the analysis of the code execution reasoning results. 

\input{listing/example_loop_iteration}

Dynamic properties should be extracted per each existing program input. To that end, \name executes the programs using a given input and leverages Trace \footnote{https://docs.python.org/3/library/trace.html} library to collect the required dynamic values such as loop lengths or recursion length.

\subsubsection{Variable Types}
\label{sec:types}
The type of variables can impact the ability of the LLMs to reason about the variable values. Intuitively, predicting the value of primitive types (e.g., integers) is likely easier than compound types (e.g., lists) and complex developer-written types (e.g., objects with multiple fields). As a result, it is necessary to consider such information when assessing LLM's abilities in code execution reasoning. \name currently considers seven variable type categories observed in the studied benchmarks, namely \texttt{\small{Int}} (e.g., $2$), \texttt{\small{Decimal}}(e.g., $2.34$), \texttt{\small{String}} (e.g., "ExeRScope"), \texttt{\small{Binary}} (e.g., True or False), \texttt{\small{List}} (e.g., [1,3,4,7]), \texttt{\small{Tuple}} (e.g., (2,7)) and \texttt{\small{Object}} (e.g., {"age":10, "gender":"female"}). 

\subsection{Visualizer}

After extracting the required properties from programs, \name aggregates the properties if needed and visualizes the results in tables or figures. \name utilizes Matplotlib~\cite{matplotlib}, pygraphviz~\cite{pygraphvis}, and networkx~\cite{networks} libraries for visualization. In addition to the visualization of results, \name is equipped with a series of statistical analysis commands to help users draw meaningful conclusions based on observations. All the data generated by the Visualizer component is aggregated into a detailed report to the end users.

%% file: listing/example_loop_iteration.tex
\begin{lstlisting}[style=stylewithcommentPy, caption={A Python code snippet with recursive logic}, label={lst:python-recursive}]
n = int(input())
l = int(input())
diff = []
for i in range(1, n + 1):
    value = l + i - 1
    if value < 0:
        value = -1 * value
    diff.append(value)
min_index = diff.index(min(diff))
print(min_index)
\end{lstlisting}

%% file: sections/evaluation.tex
\section{\name in Practice}

To demonstrate the practicality of \name and its usefulness in providing additional insights concerning code execution reasoning, we analyzed the output prediction results of six LLMs\footnote{We have covered all the family models studied by prior work. From family, we chose the biggest studied size, as they outperformed smaller models in the family.} (\gptf~\cite{chatgpt4}, \gemini~\cite{team2023gemini}, \codellama~(Instruct-34b)~\cite{roziere2023code}, \deepseek~(Instruct-33b)\cite{bi2024deepseek}, \sem(6.7b)\cite{ding2024semcoder}, and \starT~(15b)\cite{lozhkov2024starcoder}) on four benchmarks used by existing code execution reasoning frameworks: Avatar~\cite{ahmad2021avatar}, ClassEval~\cite{du2024evaluating}, CRUXEval~\cite{gu2024cruxeval}, and HumanEval~\cite{humaneval}. For a given LLM, when the programs overlap between any of the three studied frameworks, we chose the correct prediction if it existed. This empirical design is motivated by the fact that some frameworks perform better prompt engineering and help \name to focus on the real impact of program properties in code execution reasoning.   

\begin{figure}
    \centering
    \includegraphics[width=0.90\linewidth]{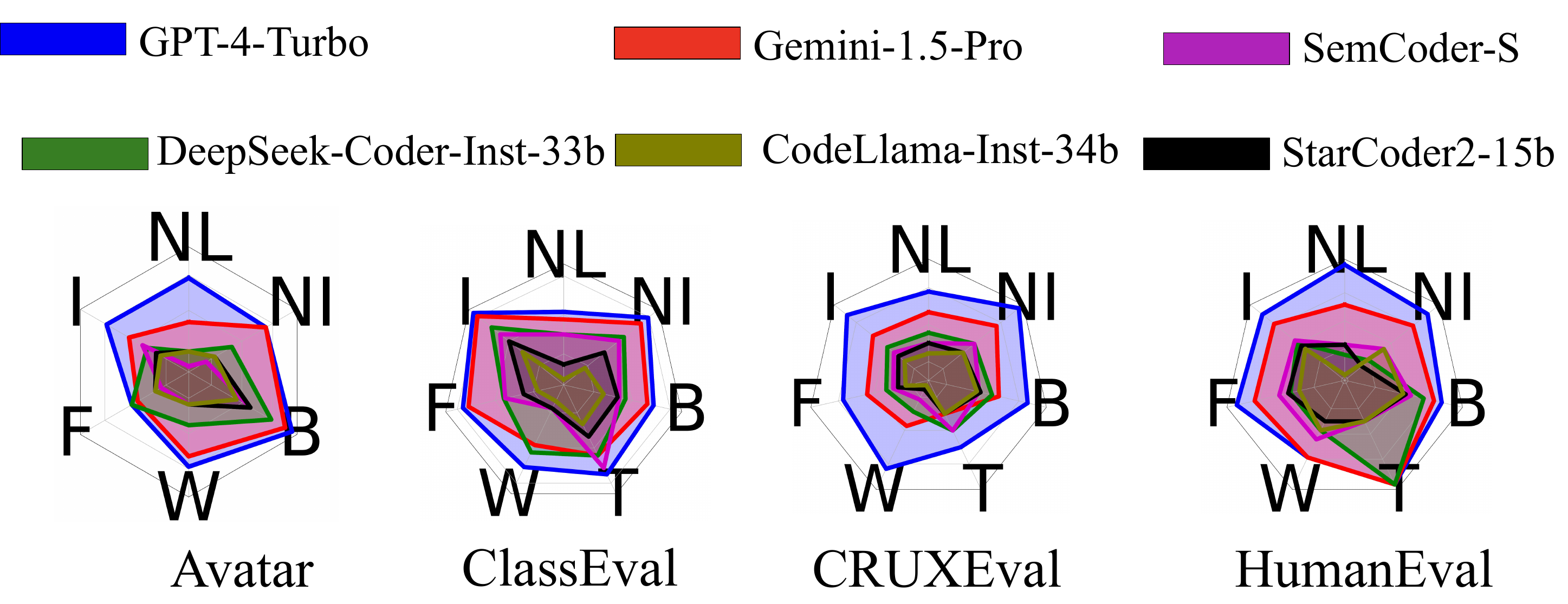}
    \caption{Impact of different programming constructs on code execution reasoning of LLMs. We abbreviate the tags 
    with B (Basic), F (For loop), I (If statement), NI (Nested Ifs), NL (Nested Loops), T (Try), and W (While loop).}
    \label{fig:construct}
    \vspace{-15pt}
\end{figure}

\subsection{Impact of Program Constructs}
To investigate the impact of different programming constructs on LLM's abilities in code reasoning, \name clusters all the input programs to the tool per their tags (\S\ref{sec:construct}) and calculates the percentage of \emph{correct value predictions} per each cluster. Figure~\ref{fig:construct} shows the visualizations provided by \name for the four benchmarks in this study, from which we observe that LLMs, especially smaller open-source ones, handle conditional statements better than recursions (while loops and for loops). Meanwhile, they struggle to reason about code execution under nested constructs such as nested if statements or nested loops.

\subsection{Impact of Program Complexity}
\begin{figure}
    \centering
    \includegraphics[width=0.99\linewidth]{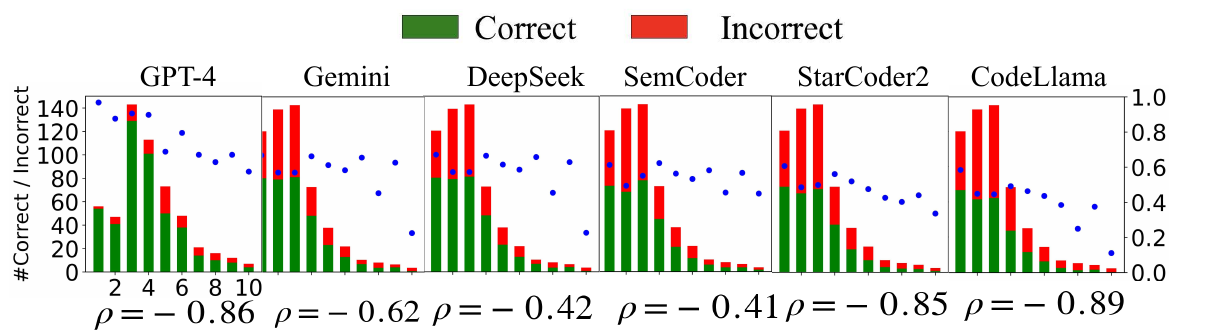}
    \caption{Impact of cyclomatic complexity on execution reasoning of LLMs. The $\rho$ symbol denotes the Spearman's Rank Order Correlation (ROC) coefficient. The blue dots demonstrate the percentage of correct variable prediction.}
    \label{fig:roc-cc}
    \vspace{-10pt}
\end{figure}

Intuitively and as discussed before (\S\ref{subsec:complexity}), a more complex program is more challenging for LLMs concerning execution reasoning. \name aids users with statistical analysis and visualizations to validate or refute this hypothesis for subject programs, i.e., those that LLMs are asked to reason about their execution. Specifically, \name computes the Spearman's Rank Order Correlation (ROC)~\cite{spearman1961proof}) between the measured program complexity metric of interest (lines of code, cyclomatic complexity, or cognitive complexity) and the success of LLMs in correct variable value prediction. 

Figure~\ref{fig:roc-cc} presents the test results ($\rho$ values) and visualizations provided by \name considering the programs in four studied benchmarks per each studied LLM and cyclomatic complexity as program complexity metric. We can see that there is always a moderate to strong negative correlation between cyclomatic complexity and LLM's success in correct output prediction, confirming the struggle of LLMs when reasoning about executing a more complex code. The bar charts visualize the programs clustered based on the cyclomatic complexity values, and the portions LLMs correctly (green bars) or incorrectly (red bars) predict their output values. The visualization, first of all, shows that programs are not very complex compared to real-world programs. Furthermore, the blue dots in the visualizations show a non-monotonic change in the LLM's ability in output prediction. That is, while the correlation between program complexity and reasoning abilities is negative, there seem to be more factors playing a role. As a result, the code execution reasoning frameworks may focus on factors other than program complexity when constructing evaluation benchmarks.

\subsection{Impact of Dynamic Program Properties}

\begin{figure}
    \centering
    \includegraphics[width=0.95\linewidth]{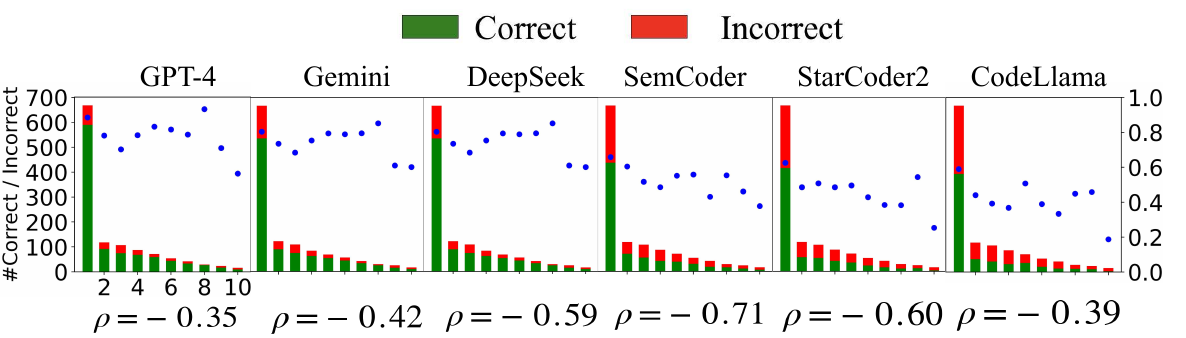}
    \caption{Impact of Loop Length on execution reasoning of LLMs. The $\rho$ symbol denotes the Spearman's Rank Order Correlation (ROC) coefficient. The blue dots demonstrate the percentage of correct variable prediction.}
    \label{fig:roc-ll}
    \vspace{-10pt}
\end{figure}

Similar to the analysis based on program complexity, \name provides statistical testing and visualization for investigating the impact of dynamic program properties on code execution reasoning. Figure~\ref{fig:roc-ll} shows the results of such analysis for \emph{loop length}, considering both for loops and while loops in the studied programs. Similar to the trend observed in Figure~\ref{fig:roc-cc}, there is consistently a moderate to strong negative correlation between loop length and LLM's performance in output prediction. This finding highlights the difficulty LLMs face in accurately tracking data flow during each loop iteration as the number of iterations increases.

\subsection{Impact of Variable Types}
\begin{figure}
    \centering
    \includegraphics[width=0.99\linewidth]{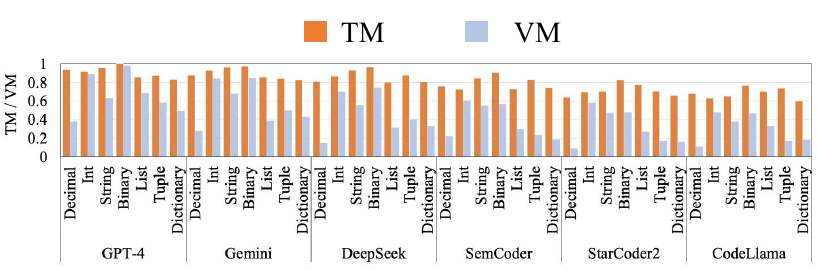}
    \caption{Performance of LLMs on predicting different variable types and values of variables. TM and VM stand for Type Match and Value Match, respectively.}
    \label{fig:type}
    \vspace{-10pt}
\end{figure}

\name aggregates the variables in the subject programs per seven categories of variable types (\S \ref{sec:types}) and checks whether an LLM correctly predicted \textit{type} of variables (Type Match) and (2) \textit{value} of variables (Value Match). Figure~\ref{fig:type} shows the visualization provided by \name for such analysis. We can observe that LLMs achieve a high Type Match (TM $>80\%$) among all the studied programs. However, when it comes to value prediction of correctly predicted types, they struggle. Considering type categories, LLMs are more successful at TM and VM for primitive types and struggle with more complex types such as Tuples, Lists, and Decimals. We speculate this is because complex types consist of multiple items with primitive or non-primitive types, recursively challenging LLMs for type and value match. 

\subsection{Data Availability and Tool Demo}
\name is publicly available~\cite{website} and can be plugged into any code execution reasoning framework. The ReadMe includes all the necessary steps to install the requirements, set up the tool, and reproduce the results and figures presented in this paper. We have provided a recorded demo of \name on the artifact website to facilitate usability.

%% file: sections/conslusion.tex
\section{Conclusion}
This paper presents \name, a tool utilizing static and dynamic program analysis to provide additional insights into the code execution reasoning abilities of LLMs. The current version of \name extracts program properties such as programming constructs, program complexity, dynamic program properties, and variable types. It then analyzes LLM's performance in code execution reasoning concerning these properties. \name is publicly available, and to demonstrate its practicality, we have analyzed output prediction results of $1450$ Python programs used by prior techniques to evaluate the reasoning abilities of six LLMs. From the provided analysis and visualizations generated by \name, we observe that higher code complexity, recursive and nested program constructs, longer loops, and non-primitive variable types are challenging for LLMs concerning code execution reasoning.

%% file: main.bbl

\begin{thebibliography}{39}


\ifx \showCODEN    \undefined \def \showCODEN     #1{\unskip}     \fi
\ifx \showISBNx    \undefined \def \showISBNx     #1{\unskip}     \fi
\ifx \showISBNxiii \undefined \def \showISBNxiii  #1{\unskip}     \fi
\ifx \showISSN     \undefined \def \showISSN      #1{\unskip}     \fi
\ifx \showLCCN     \undefined \def \showLCCN      #1{\unskip}     \fi
\ifx \shownote     \undefined \def \shownote      #1{#1}          \fi
\ifx \showarticletitle \undefined \def \showarticletitle #1{#1}   \fi
\ifx \showURL      \undefined \def \showURL       {\relax}        \fi
\providecommand\bibfield[2]{#2}
\providecommand\bibinfo[2]{#2}
\providecommand\natexlab[1]{#1}
\providecommand\showeprint[2][]{arXiv:#2}

\bibitem[mat(2025)]%
        {matplotlib}
 \bibinfo{year}{2025}\natexlab{}.
\newblock \bibinfo{title}{Matplot library for visualization}.
\newblock \bibinfo{howpublished}{\url{https://matplotlib.org/}}.
\newblock


\bibitem[net(2025)]%
        {networks}
 \bibinfo{year}{2025}\natexlab{}.
\newblock \bibinfo{title}{Network Analysis in Python using NetworkX}.
\newblock \bibinfo{howpublished}{\url{https://networkx.org/}}.
\newblock


\bibitem[pyg(2025)]%
        {pygraphvis}
 \bibinfo{year}{2025}\natexlab{}.
\newblock \bibinfo{title}{PyGraphVis library for visualization}.
\newblock \bibinfo{howpublished}{\url{https://pygraphviz.github.io/}}.
\newblock


\bibitem[ast(2025)]%
        {ast}
 \bibinfo{year}{2025}\natexlab{}.
\newblock \bibinfo{title}{Python AST module}.
\newblock \bibinfo{howpublished}{\url{https://docs.python.org/3/library/ast.html}}.
\newblock


\bibitem[Achiam et~al\mbox{.}(2023)]%
        {achiam2023gpt}
\bibfield{author}{\bibinfo{person}{Josh Achiam}, \bibinfo{person}{Steven Adler}, \bibinfo{person}{Sandhini Agarwal}, \bibinfo{person}{Lama Ahmad}, \bibinfo{person}{Ilge Akkaya}, \bibinfo{person}{Florencia~Leoni Aleman}, \bibinfo{person}{Diogo Almeida}, \bibinfo{person}{Janko Altenschmidt}, \bibinfo{person}{Sam Altman}, \bibinfo{person}{Shyamal Anadkat}, {et~al\mbox{.}}} \bibinfo{year}{2023}\natexlab{}.
\newblock \showarticletitle{Gpt-4 technical report}.
\newblock \bibinfo{journal}{\emph{arXiv preprint arXiv:2303.08774}} (\bibinfo{year}{2023}).
\newblock


\bibitem[Ahmad et~al\mbox{.}(2021)]%
        {ahmad2021avatar}
\bibfield{author}{\bibinfo{person}{Wasi~Uddin Ahmad}, \bibinfo{person}{Md~Golam~Rahman Tushar}, \bibinfo{person}{Saikat Chakraborty}, {and} \bibinfo{person}{Kai-Wei Chang}.} \bibinfo{year}{2021}\natexlab{}.
\newblock \showarticletitle{Avatar: A parallel corpus for java-python program translation}.
\newblock \bibinfo{journal}{\emph{arXiv preprint arXiv:2108.11590}} (\bibinfo{year}{2021}).
\newblock


\bibitem[Bi et~al\mbox{.}(2024)]%
        {bi2024deepseek}
\bibfield{author}{\bibinfo{person}{Xiao Bi}, \bibinfo{person}{Deli Chen}, \bibinfo{person}{Guanting Chen}, \bibinfo{person}{Shanhuang Chen}, \bibinfo{person}{Damai Dai}, \bibinfo{person}{Chengqi Deng}, \bibinfo{person}{Honghui Ding}, \bibinfo{person}{Kai Dong}, \bibinfo{person}{Qiushi Du}, \bibinfo{person}{Zhe Fu}, {et~al\mbox{.}}} \bibinfo{year}{2024}\natexlab{}.
\newblock \showarticletitle{DeepSeek LLM: Scaling Open-Source Language Models with Longtermism}.
\newblock \bibinfo{journal}{\emph{arXiv preprint arXiv:2401.02954}} (\bibinfo{year}{2024}).
\newblock


\bibitem[Bubeck et~al\mbox{.}(2023)]%
        {bubeck2023sparks}
\bibfield{author}{\bibinfo{person}{S{\'e}bastien Bubeck}, \bibinfo{person}{Varun Chandrasekaran}, \bibinfo{person}{Ronen Eldan}, \bibinfo{person}{Johannes Gehrke}, \bibinfo{person}{Eric Horvitz}, \bibinfo{person}{Ece Kamar}, \bibinfo{person}{Peter Lee}, \bibinfo{person}{Yin~Tat Lee}, \bibinfo{person}{Yuanzhi Li}, \bibinfo{person}{Scott Lundberg}, {et~al\mbox{.}}} \bibinfo{year}{2023}\natexlab{}.
\newblock \showarticletitle{Sparks of artificial general intelligence: Early experiments with gpt-4}.
\newblock \bibinfo{journal}{\emph{arXiv preprint arXiv:2303.12712}} (\bibinfo{year}{2023}).
\newblock


\bibitem[Chen et~al\mbox{.}(2024)]%
        {chen2024reasoning}
\bibfield{author}{\bibinfo{person}{Junkai Chen}, \bibinfo{person}{Zhiyuan Pan}, \bibinfo{person}{Xing Hu}, \bibinfo{person}{Zhenhao Li}, \bibinfo{person}{Ge Li}, {and} \bibinfo{person}{Xin Xia}.} \bibinfo{year}{2024}\natexlab{}.
\newblock \showarticletitle{Reasoning Runtime Behavior of a Program with LLM: How Far Are We?}. In \bibinfo{booktitle}{\emph{2025 IEEE/ACM 47th International Conference on Software Engineering (ICSE)}}. IEEE Computer Society, \bibinfo{pages}{140--152}.
\newblock


\bibitem[Chen et~al\mbox{.}(2021)]%
        {humaneval}
\bibfield{author}{\bibinfo{person}{Mark Chen}, \bibinfo{person}{Jerry Tworek}, \bibinfo{person}{Heewoo Jun}, \bibinfo{person}{Qiming Yuan}, \bibinfo{person}{Henrique~Ponde de Oliveira~Pinto}, \bibinfo{person}{Jared Kaplan}, \bibinfo{person}{Harri Edwards}, \bibinfo{person}{Yuri Burda}, \bibinfo{person}{Nicholas Joseph}, \bibinfo{person}{Greg Brockman}, \bibinfo{person}{Alex Ray}, \bibinfo{person}{Raul Puri}, \bibinfo{person}{Gretchen Krueger}, \bibinfo{person}{Michael Petrov}, \bibinfo{person}{Heidy Khlaaf}, \bibinfo{person}{Girish Sastry}, \bibinfo{person}{Pamela Mishkin}, \bibinfo{person}{Brooke Chan}, \bibinfo{person}{Scott Gray}, \bibinfo{person}{Nick Ryder}, \bibinfo{person}{Mikhail Pavlov}, \bibinfo{person}{Alethea Power}, \bibinfo{person}{Lukasz Kaiser}, \bibinfo{person}{Mohammad Bavarian}, \bibinfo{person}{Clemens Winter}, \bibinfo{person}{Philippe Tillet}, \bibinfo{person}{Felipe~Petroski Such}, \bibinfo{person}{Dave Cummings}, \bibinfo{person}{Matthias Plappert}, \bibinfo{person}{Fotios Chantzis},
  \bibinfo{person}{Elizabeth Barnes}, \bibinfo{person}{Ariel Herbert-Voss}, \bibinfo{person}{William~Hebgen Guss}, \bibinfo{person}{Alex Nichol}, \bibinfo{person}{Alex Paino}, \bibinfo{person}{Nikolas Tezak}, \bibinfo{person}{Jie Tang}, \bibinfo{person}{Igor Babuschkin}, \bibinfo{person}{Suchir Balaji}, \bibinfo{person}{Shantanu Jain}, \bibinfo{person}{William Saunders}, \bibinfo{person}{Christopher Hesse}, \bibinfo{person}{Andrew~N. Carr}, \bibinfo{person}{Jan Leike}, \bibinfo{person}{Josh Achiam}, \bibinfo{person}{Vedant Misra}, \bibinfo{person}{Evan Morikawa}, \bibinfo{person}{Alec Radford}, \bibinfo{person}{Matthew Knight}, \bibinfo{person}{Miles Brundage}, \bibinfo{person}{Mira Murati}, \bibinfo{person}{Katie Mayer}, \bibinfo{person}{Peter Welinder}, \bibinfo{person}{Bob McGrew}, \bibinfo{person}{Dario Amodei}, \bibinfo{person}{Sam McCandlish}, \bibinfo{person}{Ilya Sutskever}, {and} \bibinfo{person}{Wojciech Zaremba}.} \bibinfo{year}{2021}\natexlab{}.
\newblock \bibinfo{title}{Evaluating Large Language Models Trained on Code}.
\newblock
\showeprint[arxiv]{2107.03374}~[cs.LG]


\bibitem[Deshpande et~al\mbox{.}(2021)]%
        {deshpande2021rect}
\bibfield{author}{\bibinfo{person}{Rohan Deshpande}, \bibinfo{person}{Jerry Chen}, {and} \bibinfo{person}{Isabelle Lee}.} \bibinfo{year}{2021}\natexlab{}.
\newblock \showarticletitle{RecT: A Recursive Transformer Architecture for Generalizable Mathematical Reasoning.}. In \bibinfo{booktitle}{\emph{NeSy}}. \bibinfo{pages}{165--175}.
\newblock


\bibitem[Ding et~al\mbox{.}(2024)]%
        {ding2024semcoder}
\bibfield{author}{\bibinfo{person}{Yangruibo Ding}, \bibinfo{person}{Jinjun Peng}, \bibinfo{person}{Marcus~J Min}, \bibinfo{person}{Gail Kaiser}, \bibinfo{person}{Junfeng Yang}, {and} \bibinfo{person}{Baishakhi Ray}.} \bibinfo{year}{2024}\natexlab{}.
\newblock \showarticletitle{SemCoder: Training Code Language Models with Comprehensive Semantics}.
\newblock \bibinfo{journal}{\emph{arXiv preprint arXiv:2406.01006}} (\bibinfo{year}{2024}).
\newblock


\bibitem[Du et~al\mbox{.}(2024)]%
        {du2024evaluating}
\bibfield{author}{\bibinfo{person}{Xueying Du}, \bibinfo{person}{Mingwei Liu}, \bibinfo{person}{Kaixin Wang}, \bibinfo{person}{Hanlin Wang}, \bibinfo{person}{Junwei Liu}, \bibinfo{person}{Yixuan Chen}, \bibinfo{person}{Jiayi Feng}, \bibinfo{person}{Chaofeng Sha}, \bibinfo{person}{Xin Peng}, {and} \bibinfo{person}{Yiling Lou}.} \bibinfo{year}{2024}\natexlab{}.
\newblock \showarticletitle{Evaluating large language models in class-level code generation}. In \bibinfo{booktitle}{\emph{Proceedings of the IEEE/ACM 46th International Conference on Software Engineering}}. \bibinfo{pages}{1--13}.
\newblock


\bibitem[Gu et~al\mbox{.}(2024)]%
        {gu2024cruxeval}
\bibfield{author}{\bibinfo{person}{Alex Gu}, \bibinfo{person}{Baptiste Rozi{\`e}re}, \bibinfo{person}{Hugh Leather}, \bibinfo{person}{Armando Solar-Lezama}, \bibinfo{person}{Gabriel Synnaeve}, {and} \bibinfo{person}{Sida~I Wang}.} \bibinfo{year}{2024}\natexlab{}.
\newblock \showarticletitle{CRUXEval: A Benchmark for Code Reasoning, Understanding and Execution}.
\newblock \bibinfo{journal}{\emph{arXiv preprint arXiv:2401.03065}} (\bibinfo{year}{2024}).
\newblock


\bibitem[Huang et~al\mbox{.}(2023)]%
        {huang2023lvlms}
\bibfield{author}{\bibinfo{person}{Kung-Hsiang Huang}, \bibinfo{person}{Mingyang Zhou}, \bibinfo{person}{Hou~Pong Chan}, \bibinfo{person}{Yi~R Fung}, \bibinfo{person}{Zhenhailong Wang}, \bibinfo{person}{Lingyu Zhang}, \bibinfo{person}{Shih-Fu Chang}, {and} \bibinfo{person}{Heng Ji}.} \bibinfo{year}{2023}\natexlab{}.
\newblock \showarticletitle{Do LVLMs Understand Charts? Analyzing and Correcting Factual Errors in Chart Captioning}.
\newblock \bibinfo{journal}{\emph{arXiv preprint arXiv:2312.10160}} (\bibinfo{year}{2023}).
\newblock


\bibitem[Imani et~al\mbox{.}(2023)]%
        {imani2023mathprompter}
\bibfield{author}{\bibinfo{person}{Shima Imani}, \bibinfo{person}{Liang Du}, {and} \bibinfo{person}{Harsh Shrivastava}.} \bibinfo{year}{2023}\natexlab{}.
\newblock \showarticletitle{Mathprompter: Mathematical reasoning using large language models}.
\newblock \bibinfo{journal}{\emph{arXiv preprint arXiv:2303.05398}} (\bibinfo{year}{2023}).
\newblock


\bibitem[Kang et~al\mbox{.}(2024)]%
        {kang2024learning}
\bibfield{author}{\bibinfo{person}{Katie Kang}, \bibinfo{person}{Amrith Setlur}, \bibinfo{person}{Dibya Ghosh}, \bibinfo{person}{Jacob Steinhardt}, \bibinfo{person}{Claire Tomlin}, \bibinfo{person}{Sergey Levine}, {and} \bibinfo{person}{Aviral Kumar}.} \bibinfo{year}{2024}\natexlab{}.
\newblock \showarticletitle{What Do Learning Dynamics Reveal About Generalization in LLM Reasoning?}
\newblock \bibinfo{journal}{\emph{arXiv preprint arXiv:2411.07681}} (\bibinfo{year}{2024}).
\newblock


\bibitem[La~Malfa et~al\mbox{.}(2024)]%
        {la2024code}
\bibfield{author}{\bibinfo{person}{Emanuele La~Malfa}, \bibinfo{person}{Christoph Weinhuber}, \bibinfo{person}{Orazio Torre}, \bibinfo{person}{Fangru Lin}, \bibinfo{person}{Anthony Cohn}, \bibinfo{person}{Nigel Shadbolt}, {and} \bibinfo{person}{Michael Wooldridge}.} \bibinfo{year}{2024}\natexlab{}.
\newblock \showarticletitle{Code Simulation Challenges for Large Language Models}.
\newblock \bibinfo{journal}{\emph{arXiv preprint arXiv:2401.09074}} (\bibinfo{year}{2024}).
\newblock


\bibitem[Liu et~al\mbox{.}(2023)]%
        {liu2023code}
\bibfield{author}{\bibinfo{person}{Chenxiao Liu}, \bibinfo{person}{Shuai Lu}, \bibinfo{person}{Weizhu Chen}, \bibinfo{person}{Daxin Jiang}, \bibinfo{person}{Alexey Svyatkovskiy}, \bibinfo{person}{Shengyu Fu}, \bibinfo{person}{Neel Sundaresan}, {and} \bibinfo{person}{Nan Duan}.} \bibinfo{year}{2023}\natexlab{}.
\newblock \showarticletitle{Code Execution with Pre-trained Language Models}.
\newblock \bibinfo{journal}{\emph{arXiv preprint arXiv:2305.05383}} (\bibinfo{year}{2023}).
\newblock


\bibitem[Lozhkov et~al\mbox{.}(2024)]%
        {lozhkov2024starcoder}
\bibfield{author}{\bibinfo{person}{Anton Lozhkov}, \bibinfo{person}{Raymond Li}, \bibinfo{person}{Loubna~Ben Allal}, \bibinfo{person}{Federico Cassano}, \bibinfo{person}{Joel Lamy-Poirier}, \bibinfo{person}{Nouamane Tazi}, \bibinfo{person}{Ao Tang}, \bibinfo{person}{Dmytro Pykhtar}, \bibinfo{person}{Jiawei Liu}, \bibinfo{person}{Yuxiang Wei}, {et~al\mbox{.}}} \bibinfo{year}{2024}\natexlab{}.
\newblock \showarticletitle{StarCoder 2 and The Stack v2: The Next Generation}.
\newblock \bibinfo{journal}{\emph{arXiv preprint arXiv:2402.19173}} (\bibinfo{year}{2024}).
\newblock


\bibitem[Luo et~al\mbox{.}(2023)]%
        {luo2023wizardmath}
\bibfield{author}{\bibinfo{person}{Haipeng Luo}, \bibinfo{person}{Qingfeng Sun}, \bibinfo{person}{Can Xu}, \bibinfo{person}{Pu Zhao}, \bibinfo{person}{Jianguang Lou}, \bibinfo{person}{Chongyang Tao}, \bibinfo{person}{Xiubo Geng}, \bibinfo{person}{Qingwei Lin}, \bibinfo{person}{Shifeng Chen}, {and} \bibinfo{person}{Dongmei Zhang}.} \bibinfo{year}{2023}\natexlab{}.
\newblock \showarticletitle{Wizardmath: Empowering mathematical reasoning for large language models via reinforced evol-instruct}.
\newblock \bibinfo{journal}{\emph{arXiv preprint arXiv:2308.09583}} (\bibinfo{year}{2023}).
\newblock


\bibitem[Miceli-Barone et~al\mbox{.}(2023)]%
        {miceli2023larger}
\bibfield{author}{\bibinfo{person}{Antonio~Valerio Miceli-Barone}, \bibinfo{person}{Fazl Barez}, \bibinfo{person}{Ioannis Konstas}, {and} \bibinfo{person}{Shay~B Cohen}.} \bibinfo{year}{2023}\natexlab{}.
\newblock \showarticletitle{The Larger They Are, the Harder They Fail: Language Models do not Recognize Identifier Swaps in Python}.
\newblock \bibinfo{journal}{\emph{arXiv preprint arXiv:2305.15507}} (\bibinfo{year}{2023}).
\newblock


\bibitem[Min et~al\mbox{.}(2023)]%
        {min2023beyond}
\bibfield{author}{\bibinfo{person}{Marcus~J Min}, \bibinfo{person}{Yangruibo Ding}, \bibinfo{person}{Luca Buratti}, \bibinfo{person}{Saurabh Pujar}, \bibinfo{person}{Gail Kaiser}, \bibinfo{person}{Suman Jana}, {and} \bibinfo{person}{Baishakhi Ray}.} \bibinfo{year}{2023}\natexlab{}.
\newblock \showarticletitle{Beyond Accuracy: Evaluating Self-Consistency of Code Large Language Models with IdentityChain}.
\newblock \bibinfo{journal}{\emph{arXiv preprint arXiv:2310.14053}} (\bibinfo{year}{2023}).
\newblock


\bibitem[Mishra et~al\mbox{.}(2024)]%
        {mishra2024granite}
\bibfield{author}{\bibinfo{person}{Mayank Mishra}, \bibinfo{person}{Matt Stallone}, \bibinfo{person}{Gaoyuan Zhang}, \bibinfo{person}{Yikang Shen}, \bibinfo{person}{Aditya Prasad}, \bibinfo{person}{Adriana~Meza Soria}, \bibinfo{person}{Michele Merler}, \bibinfo{person}{Parameswaran Selvam}, \bibinfo{person}{Saptha Surendran}, \bibinfo{person}{Shivdeep Singh}, {et~al\mbox{.}}} \bibinfo{year}{2024}\natexlab{}.
\newblock \showarticletitle{Granite code models: A family of open foundation models for code intelligence}.
\newblock \bibinfo{journal}{\emph{arXiv preprint arXiv:2405.04324}} (\bibinfo{year}{2024}).
\newblock


\bibitem[Ni et~al\mbox{.}(2024)]%
        {ni2024next}
\bibfield{author}{\bibinfo{person}{Ansong Ni}, \bibinfo{person}{Miltiadis Allamanis}, \bibinfo{person}{Arman Cohan}, \bibinfo{person}{Yinlin Deng}, \bibinfo{person}{Kensen Shi}, \bibinfo{person}{Charles Sutton}, {and} \bibinfo{person}{Pengcheng Yin}.} \bibinfo{year}{2024}\natexlab{}.
\newblock \showarticletitle{NExT: Teaching Large Language Models to Reason about Code Execution}.
\newblock \bibinfo{journal}{\emph{arXiv preprint arXiv:2404.14662}} (\bibinfo{year}{2024}).
\newblock


\bibitem[OpenAI(2023)]%
        {chatgpt4}
\bibfield{author}{\bibinfo{person}{OpenAI}.} \bibinfo{year}{2023}\natexlab{}.
\newblock \showarticletitle{GPT-4 Technical Report}.
\newblock \bibinfo{journal}{\emph{https://arxiv.org/abs/2303.08774}} (\bibinfo{year}{2023}).
\newblock


\bibitem[Reid et~al\mbox{.}(2024)]%
        {reid2024gemini}
\bibfield{author}{\bibinfo{person}{Machel Reid}, \bibinfo{person}{Nikolay Savinov}, \bibinfo{person}{Denis Teplyashin}, \bibinfo{person}{Dmitry Lepikhin}, \bibinfo{person}{Timothy Lillicrap}, \bibinfo{person}{Jean-baptiste Alayrac}, \bibinfo{person}{Radu Soricut}, \bibinfo{person}{Angeliki Lazaridou}, \bibinfo{person}{Orhan Firat}, \bibinfo{person}{Julian Schrittwieser}, {et~al\mbox{.}}} \bibinfo{year}{2024}\natexlab{}.
\newblock \showarticletitle{Gemini 1.5: Unlocking multimodal understanding across millions of tokens of context}.
\newblock \bibinfo{journal}{\emph{arXiv preprint arXiv:2403.05530}} (\bibinfo{year}{2024}).
\newblock


\bibitem[Roziere et~al\mbox{.}(2023)]%
        {roziere2023code}
\bibfield{author}{\bibinfo{person}{Baptiste Roziere}, \bibinfo{person}{Jonas Gehring}, \bibinfo{person}{Fabian Gloeckle}, \bibinfo{person}{Sten Sootla}, \bibinfo{person}{Itai Gat}, \bibinfo{person}{Xiaoqing~Ellen Tan}, \bibinfo{person}{Yossi Adi}, \bibinfo{person}{Jingyu Liu}, \bibinfo{person}{Tal Remez}, \bibinfo{person}{J{\'e}r{\'e}my Rapin}, {et~al\mbox{.}}} \bibinfo{year}{2023}\natexlab{}.
\newblock \showarticletitle{Code llama: Open foundation models for code}.
\newblock \bibinfo{journal}{\emph{arXiv preprint arXiv:2308.12950}} (\bibinfo{year}{2023}).
\newblock


\bibitem[Spearman(1961)]%
        {spearman1961proof}
\bibfield{author}{\bibinfo{person}{Charles Spearman}.} \bibinfo{year}{1961}\natexlab{}.
\newblock \showarticletitle{The proof and measurement of association between two things.}
\newblock  (\bibinfo{year}{1961}).
\newblock


\bibitem[Team et~al\mbox{.}(2023)]%
        {team2023gemini}
\bibfield{author}{\bibinfo{person}{Gemini Team}, \bibinfo{person}{Rohan Anil}, \bibinfo{person}{Sebastian Borgeaud}, \bibinfo{person}{Yonghui Wu}, \bibinfo{person}{Jean-Baptiste Alayrac}, \bibinfo{person}{Jiahui Yu}, \bibinfo{person}{Radu Soricut}, \bibinfo{person}{Johan Schalkwyk}, \bibinfo{person}{Andrew~M Dai}, \bibinfo{person}{Anja Hauth}, {et~al\mbox{.}}} \bibinfo{year}{2023}\natexlab{}.
\newblock \showarticletitle{Gemini: a family of highly capable multimodal models}.
\newblock \bibinfo{journal}{\emph{arXiv preprint arXiv:2312.11805}} (\bibinfo{year}{2023}).
\newblock


\bibitem[Valmeekam et~al\mbox{.}(2022)]%
        {valmeekam2022large}
\bibfield{author}{\bibinfo{person}{Karthik Valmeekam}, \bibinfo{person}{Alberto Olmo}, \bibinfo{person}{Sarath Sreedharan}, {and} \bibinfo{person}{Subbarao Kambhampati}.} \bibinfo{year}{2022}\natexlab{}.
\newblock \showarticletitle{Large Language Models Still Can't Plan (A Benchmark for LLMs on Planning and Reasoning about Change)}.
\newblock \bibinfo{journal}{\emph{arXiv preprint arXiv:2206.10498}} (\bibinfo{year}{2022}).
\newblock


\bibitem[Wang et~al\mbox{.}(2023)]%
        {wang2023mathcoder}
\bibfield{author}{\bibinfo{person}{Ke Wang}, \bibinfo{person}{Houxing Ren}, \bibinfo{person}{Aojun Zhou}, \bibinfo{person}{Zimu Lu}, \bibinfo{person}{Sichun Luo}, \bibinfo{person}{Weikang Shi}, \bibinfo{person}{Renrui Zhang}, \bibinfo{person}{Linqi Song}, \bibinfo{person}{Mingjie Zhan}, {and} \bibinfo{person}{Hongsheng Li}.} \bibinfo{year}{2023}\natexlab{}.
\newblock \showarticletitle{Mathcoder: Seamless code integration in llms for enhanced mathematical reasoning}.
\newblock \bibinfo{journal}{\emph{arXiv preprint arXiv:2310.03731}} (\bibinfo{year}{2023}).
\newblock


\bibitem[Website(2025)]%
        {website}
\bibfield{author}{\bibinfo{person}{ExeRScope Tool~Artifact Website}.} \bibinfo{year}{2025}\natexlab{}.
\newblock \bibinfo{howpublished}{\url{https://github.com/Intelligent-CAT-Lab/ExeRScope}}.
\newblock


\bibitem[Wu et~al\mbox{.}(2023)]%
        {wu2023reasoning}
\bibfield{author}{\bibinfo{person}{Zhaofeng Wu}, \bibinfo{person}{Linlu Qiu}, \bibinfo{person}{Alexis Ross}, \bibinfo{person}{Ekin Aky{\"u}rek}, \bibinfo{person}{Boyuan Chen}, \bibinfo{person}{Bailin Wang}, \bibinfo{person}{Najoung Kim}, \bibinfo{person}{Jacob Andreas}, {and} \bibinfo{person}{Yoon Kim}.} \bibinfo{year}{2023}\natexlab{}.
\newblock \showarticletitle{Reasoning or reciting? exploring the capabilities and limitations of language models through counterfactual tasks}.
\newblock \bibinfo{journal}{\emph{arXiv preprint arXiv:2307.02477}} (\bibinfo{year}{2023}).
\newblock


\bibitem[Yang et~al\mbox{.}(2024)]%
        {yang2024qwen2}
\bibfield{author}{\bibinfo{person}{An Yang}, \bibinfo{person}{Baosong Yang}, \bibinfo{person}{Binyuan Hui}, \bibinfo{person}{Bo Zheng}, \bibinfo{person}{Bowen Yu}, \bibinfo{person}{Chang Zhou}, \bibinfo{person}{Chengpeng Li}, \bibinfo{person}{Chengyuan Li}, \bibinfo{person}{Dayiheng Liu}, \bibinfo{person}{Fei Huang}, {et~al\mbox{.}}} \bibinfo{year}{2024}\natexlab{}.
\newblock \showarticletitle{Qwen2 technical report}.
\newblock \bibinfo{journal}{\emph{arXiv preprint arXiv:2407.10671}} (\bibinfo{year}{2024}).
\newblock


\bibitem[Yin(2025)]%
        {lizard}
\bibfield{author}{\bibinfo{person}{Terry Yin}.} \bibinfo{year}{2025}\natexlab{}.
\newblock \bibinfo{title}{Lizard}.
\newblock \bibinfo{howpublished}{\url{https://github.com/terryyin/lizard}}.
\newblock


\bibitem[Zhang et~al\mbox{.}(2024a)]%
        {zhang2024transformer}
\bibfield{author}{\bibinfo{person}{Dylan Zhang}, \bibinfo{person}{Curt Tigges}, \bibinfo{person}{Zory Zhang}, \bibinfo{person}{Stella Biderman}, \bibinfo{person}{Maxim Raginsky}, {and} \bibinfo{person}{Talia Ringer}.} \bibinfo{year}{2024}\natexlab{a}.
\newblock \showarticletitle{Transformer-Based Models Are Not Yet Perfect At Learning to Emulate Structural Recursion}.
\newblock \bibinfo{journal}{\emph{arXiv preprint arXiv:2401.12947}} (\bibinfo{year}{2024}).
\newblock


\bibitem[Zhang et~al\mbox{.}(2024b)]%
        {zhang2024can}
\bibfield{author}{\bibinfo{person}{Yizhuo Zhang}, \bibinfo{person}{Heng Wang}, \bibinfo{person}{Shangbin Feng}, \bibinfo{person}{Zhaoxuan Tan}, \bibinfo{person}{Xiaochuang Han}, \bibinfo{person}{Tianxing He}, {and} \bibinfo{person}{Yulia Tsvetkov}.} \bibinfo{year}{2024}\natexlab{b}.
\newblock \showarticletitle{Can LLM Graph Reasoning Generalize beyond Pattern Memorization?}
\newblock \bibinfo{journal}{\emph{arXiv preprint arXiv:2406.15992}} (\bibinfo{year}{2024}).
\newblock


\bibitem[Zhu et~al\mbox{.}(2024)]%
        {zhu2024deepseek}
\bibfield{author}{\bibinfo{person}{Qihao Zhu}, \bibinfo{person}{Daya Guo}, \bibinfo{person}{Zhihong Shao}, \bibinfo{person}{Dejian Yang}, \bibinfo{person}{Peiyi Wang}, \bibinfo{person}{Runxin Xu}, \bibinfo{person}{Y Wu}, \bibinfo{person}{Yukun Li}, \bibinfo{person}{Huazuo Gao}, \bibinfo{person}{Shirong Ma}, {et~al\mbox{.}}} \bibinfo{year}{2024}\natexlab{}.
\newblock \showarticletitle{DeepSeek-Coder-V2: Breaking the Barrier of Closed-Source Models in Code Intelligence}.
\newblock \bibinfo{journal}{\emph{arXiv preprint arXiv:2406.11931}} (\bibinfo{year}{2024}).
\newblock


\end{thebibliography}
